\documentclass{article}
\usepackage{graphicx} 
\usepackage[english]{babel}
 
\title{Analytical approaches to the study of the phase of the visibility function}
\author{S.V. Chernov}

\begin{document}

\maketitle
\begin{abstract}
In radio interferometric observations, the main source of information is the complex visibility function, which includes amplitude and phase. In this paper, the dependence of the phase of the visibility function on the base projection is investigated when used in radio interferometry with space bases up to six Earth diameters. The dependence of the phase of the visibility function on the projection of the base and direction is obtained. It is shown that for small values of the base projections, this dependence has a universal character and is consistent with the results of numerical magnetohydrodynamic models.
\end{abstract}

\section{Introduction}

Radio interferometry with space bases is one of the most promising methods for studying supermassive black holes \cite{likhachev2022}. This approach allows you to obtain images of such objects with high resolution \cite{likhachev2022,rud2022,chernov2025}. This technology is based on the use of a complex of satellite antennas located in orbits at considerable distances in space \cite{rud2022}. This arrangement significantly expands the possibilities for obtaining detailed images and accurate measurement of source parameters that would be inaccessible using ground-based radio interferometers due to atmospheric conditions and limited distances.

The key concept in radio interferometry is the visibility function, which is a quantitative characteristic of the cross—correlation of signals received by two or more antennas \cite{thompson2017}. This function is a complex quantity that consists of the amplitude and phase \cite{thompson2017}. The analysis of the visibility function allows you to obtain information about the structure of the source, investigate physical processes and refine the parameters of space objects \cite{chernov2025}.

This article discusses the mathematical formalism of the phase of the visibility function in the conditions of radio interferometry with space bases. Fairly general formulas are written out that separate the real part from the imaginary part. For small projections of the bases, a universal relationship is also derived that relates the phase of the visibility function to the source parameters. This approach will allow for a more accurate interpretation of the observational data.

The work uses geometric units in which the gravity constant and the speed of light equals one, $G=c=1$.

\section{Basic notations and definitions}

The visibility function in polar coordinates $u,\phi_u$ can be written as follows \cite{thompson2017,chernov2025a}
\begin{eqnarray}
 V(u,\phi_u)=\int\int I(r,\phi_r) e^{-2\pi iur\cos(\phi_r-\phi_u)}rdrd\phi_r,
 \label{vid}
\end{eqnarray}
where $u$ is the projection of the base in terms of wavelengths, $r,\phi_r$ are the polar coordinates of the source in the sky in the picture plane, expressed in radians, $I(r,\phi_r)$ is the brightness distribution of the source in the sky. This expression shows that the visibility function is related to the brightness distribution of the source in the sky $I(r,\phi_r)$ via the Fourier transform. It is a complex quantity and can be represented through its amplitude (modulus) $|V|$ and phase $\phi_V$
\begin{eqnarray}
V(u,\phi_u)=|V|e^{i\phi_V}=ReV+iImV,
\end{eqnarray}
where $ReV$ is the real part and $ImV$ is the imaginary part of the visibility function.

Suppose that the brightness distribution function of a source on a celestial sphere can be represented as the product of two functions: radial $I_r$ and azimuthal $I_\phi$\cite{chernov2025}. It is assumed that the radial brightness distribution function of the source depends only on the radial coordinate $r$, and the azimuthal one depends only on the azimuthal coordinate $\phi_r$. In other words, the brightness distribution function of the source can be represented as
\begin{eqnarray}
I(r,\phi_r)=I_r(r)I_\phi(\phi_r).
\label{IrphiIrIphi}
\end{eqnarray}
Then, if we use the properties of the generating function \cite{yanke1977}
\begin{eqnarray}
e^{-iz\cos\phi}=\sum_{n=-\infty}^{\infty} (-i)^ne^{in\phi}J_n(z),
\label{sumJ}
\end{eqnarray}
where $J_n$ are Bessel functions of the first kind of order $n$, then the integral expression in the visibility function (\ref{vid}) can be represented as the sum of the product of single integrals. Substituting the expressions (\ref{IrphiIrIphi}) and (\ref{sumJ}) into the formula (\ref{vid}), we obtain
\begin{eqnarray}
 V(u,\phi_u)
 =\sum_{n=-\infty}^{\infty} (-i)^n\int_0^\infty I_r(r)J_n\left(2\pi ur\right)rdr\int_{0}^{2\pi} I_\phi(\phi_r)  e^{in(\phi_r-\phi_u)}d\phi_r.
 \label{sumvid}
\end{eqnarray}
These single integrals can be easily calculated in a fairly general case.

To calculate the azimuthal integral in the expression (\ref{sumvid}), we decompose the azimuthal part of the brightness distribution function $I_\phi$ into a Fourier series over the interval $[0.2\pi]$
\begin{eqnarray}
I_\phi(\phi_r)=\frac{a_0}{2}+\sum_{m=1}^{\infty}(a_m\cos m(\phi_r-\phi_0)+b_m\sin m(\phi_r-\phi_0)),
\label{iphia}
\end{eqnarray}
where $a_m$, $b_m$ are constant Fourier coefficients, and $\phi_0$ is the direction of maximum brightness. Then the second integral in (\ref{sumvid}) can be calculated explicitly. To do this, use the following expressions:
If $I_\phi=\frac{a_0}{2}$, then 
\begin{eqnarray}
 \int_0^{2\pi}\frac{a_0}{2}e^{in(\phi_r-\phi_u)}d\phi_r=\pi\delta_{n,0}a_0,
 \label{a0}
\end{eqnarray}
where $\delta_{n,0}$ is the Kronecker symbol. 
If $I_\phi=b_m\sin(m\phi_r-m\phi_0)$, then
\begin{eqnarray}
 \int_0^{2\pi}b_m\sin(m\phi_r-m\phi_0)e^{in(\phi_r-\phi_u)}d\phi_r=\pi b_m ie^{i(\phi_0-\phi_u)n}\delta_{n,m}-\pi b_mie^{i(\phi_0-\phi_u)n}\delta_{-n,m}.
 \label{bm}
\end{eqnarray}
If $I_\phi=a_m\cos(m\phi_r-m\phi_0)$, then
\begin{eqnarray}
 \int_0^{2\pi}a_m\cos(m\phi_r-m\phi_0)e^{in(\phi_r-\phi_u)}d\phi_r=\pi a_m e^{i(\phi_0-\phi_u)n}\delta_{n,m}+\pi a_me^{i(\phi_0-\phi_u)n}\delta_{-n,m}.
 \label{am}
\end{eqnarray}
Substituting the expression (\ref{iphia}) into (\ref{sumvid}) and using (\ref{a0}), (\ref{bm}) and (\ref{am}), we get the visibility function as a result.
\begin{eqnarray}
V(u,\phi_u)=\pi a_0\int_0^{\infty}I_r(r)J_0(2\pi ur)rdr+\nonumber\\
+\sum_{m=1}^{\infty}2\pi (-i)^ma_m\cos(m(\phi_0-\phi_u))\int_0^\infty I_r(r)J_m(2\pi ur)rdr-\nonumber\\
-\sum_{m=1}^{\infty}2\pi (-i)^mb_m\sin(m(\phi_0-\phi_u))\int_0^\infty I_r(r)J_m(2\pi ur)rdr.
\label{finalvid}
\end{eqnarray}
This is the final form of the visibility function, which we will explore in this paper.
It can be noted that for odd integers $m=1,3,5\ldots$ the visibility function is complex, while for even integers $m=0,2,4\ldots$ it is real.

Consider the case of small projections of bases, when $2\pi ur\ll1$. In this case, only two terms with numbers $m=0$ and $m=1$ remain in the expression (\ref{finalvid}). The first one $m=0$ is real, and the second one $m=1$ is imaginary. As was shown in \cite{chernov2025}, for small base projection values, the first zeros of the visibility function match with good accuracy for different radial profiles of the source brightness functions. Therefore, this allows us to approximate the radial brightness of the source as a delta function: $I_r(r)\sim\delta(r-r_0)$, where $r_0$ is the radius of the ring. Substituting such an approximation into the expression for the visibility function (\ref{finalvid}), we finally get:
\begin{eqnarray}
V(u,\phi_u)=\pi a_0J_0(2\pi ur_0)r_0+\nonumber\\
-2\pi ia_1\cos(\phi_0-\phi_u)J_1(2\pi ur_0)r_0+\nonumber\\
+2\pi ib_1\sin(\phi_0-\phi_u)J_1(2\pi ur_0)r_0
\label{viddelta}
\end{eqnarray}
It follows that for small projections of the bases, the phase of the visibility function will be determined by the ratio of the first-order Bessel functions to zero.
\begin{eqnarray}
\tan\phi_v=\frac{ImV}{ReV}\sim\frac{J_1(2\pi ur_0)}{J_0(2\pi ur_0)}.
\end{eqnarray}
This dependence is universal for small projections of the bases. You can see that in the case when the numerator is zero, and this happens when \cite{zaytsev2001}
\begin{eqnarray}
 2\pi ur_0\approx\pi\left(n+\frac{1}{4}\right)-\frac{3}{8\pi(n+\frac{1}{4})}+\frac{3}{128\pi^3(n+\frac{1}{4})^3}
\end{eqnarray}
the phase of the visibility function is close to zero. In the case when the denominator is zero, that is, when \cite{zaytsev2001}
\begin{eqnarray}
 2\pi ur_0\approx\frac{\pi}{4}(4n+3)+\frac{1}{2\pi(4n+3)}-\frac{31}{6\pi^3(4n+3)^3}
\end{eqnarray}
where $n=0,1,2\ldots$ are the zeros of the Bessel function,
The phase of the visibility function reaches the values of $\pm180^\circ$. A similar dependence is characteristic for the phase of the visibility function of a binary point system with the same brightness distribution \cite{monnier2007}.

In the more general case, for arbitrary projections of bases, it is necessary to calculate radial integrals and sum all terms in the expression (\ref{finalvid}). For a more visual representation and comparison with MHD models, let us consider a special case that makes it possible to simplify calculations and easily interpret the results obtained.

\section{A special case}

\begin{figure}
\centering
\includegraphics[width=1.0\linewidth]{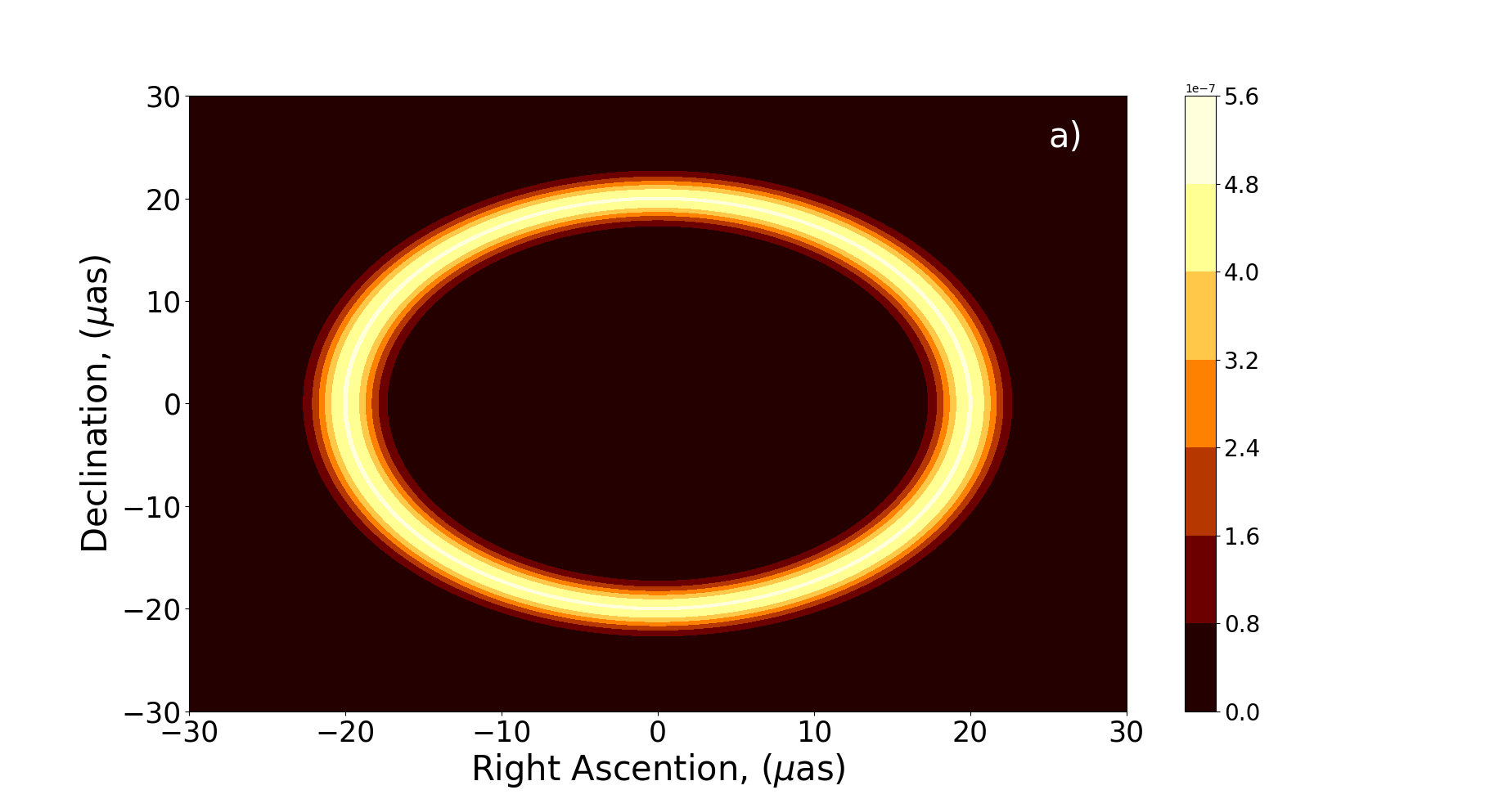}
\includegraphics[width=1.0\linewidth]{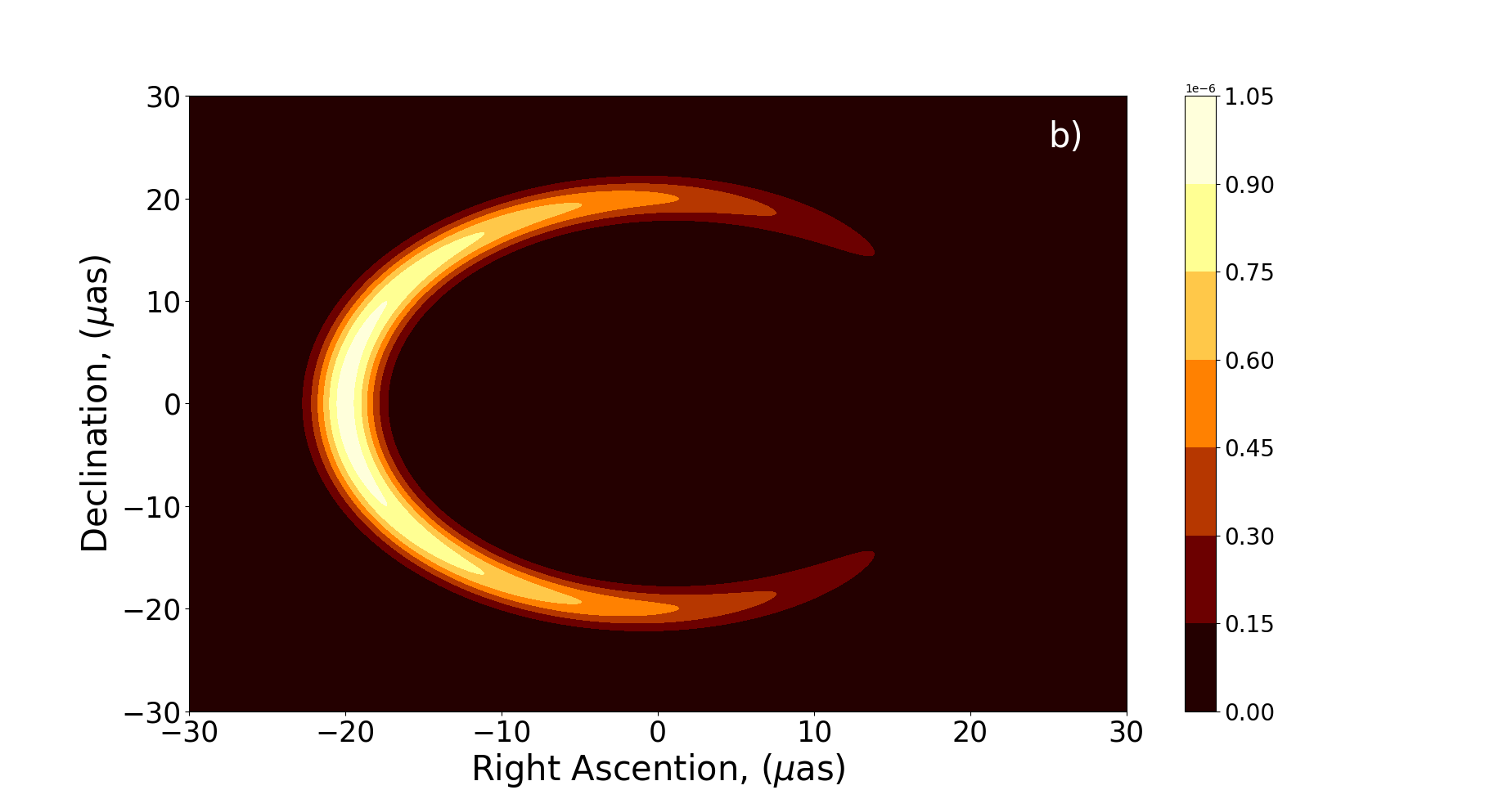}
\includegraphics[width=1.0\linewidth]{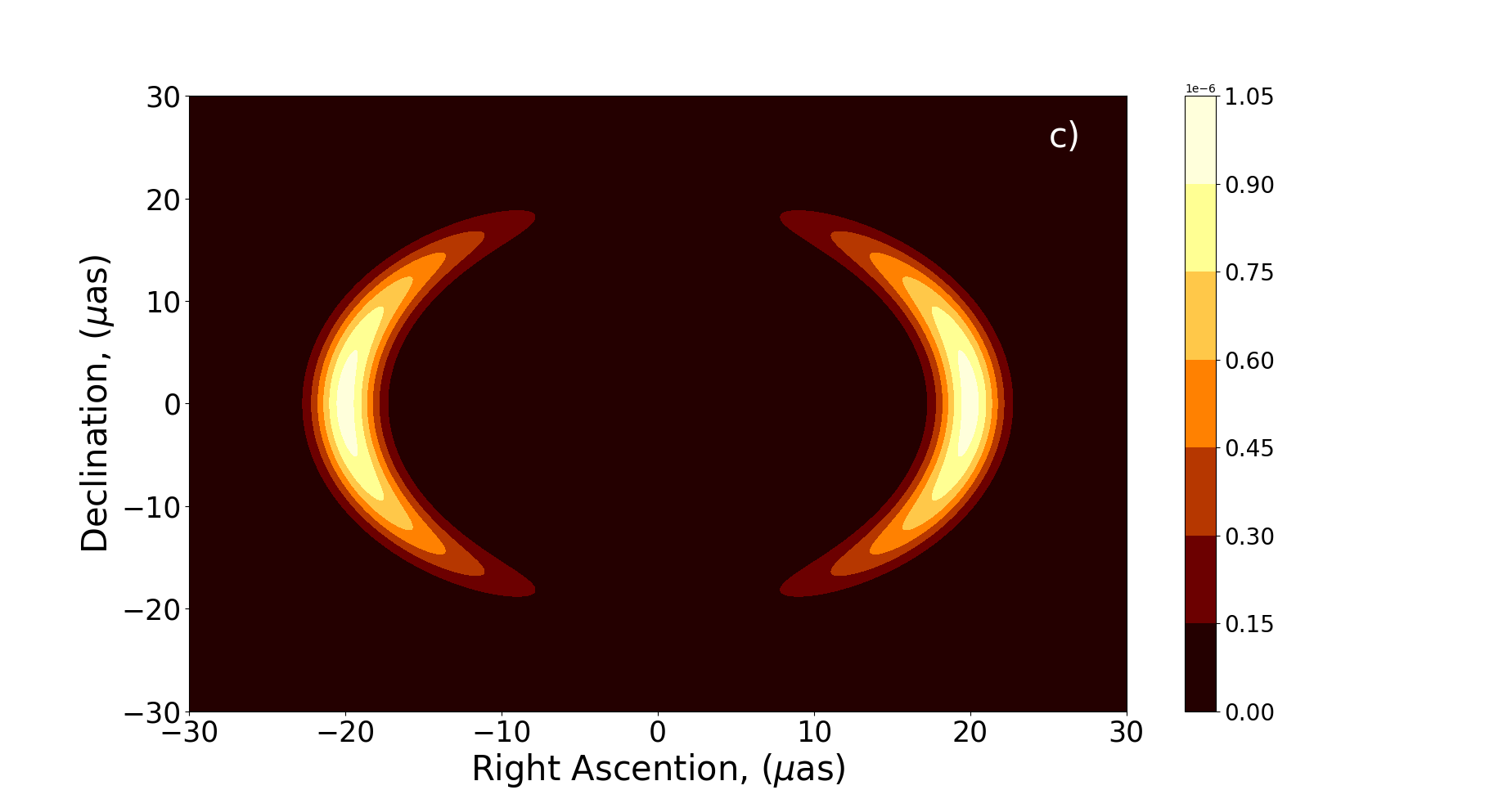}
\caption{The figure shows examples of images for cases when a) - $a_0=2$, $a_1=a_2=0$, b) - $a_0=2$, $a_1=1$, $a_2=0$ and c) - $a_0=2$, $a_1=0$, $a_2=1$.}
\label{fig1}
\end{figure}

Let's consider an example of a special case when the azimuthal brightness function is given as the sum of three terms of the Fourier series ($m=0,1,2,$ see \cite{chernov2025a})
\begin{eqnarray}
I_\phi=\frac{a_0}{2}+a_1\cos(\phi_r-\phi_0)+a_2\cos(2\phi_r-2\phi_0).
\label{iphifur}
\end{eqnarray}
The figure (\ref{fig1}) shows examples of images with sizes of $60$ per $60$ microseconds ($\mu as$) of arc for various sets of parameters a) - $a_0=2$, $a_1=a_2=0$, b) - $a_0=2$, $a_1=1$, $a_2=0$, c) - $a_0=2$, $a_1=0$, $a_2=1$. The radial brightness profile of the ring $I_r(r)$ was modeled using a Gaussian function with a ring radius of 20 microseconds of arc and a width of 2 microseconds of arc \cite{chernov2025a,chernov2026}. The direction angle for maximum brightness was set to $\phi_0=\pi$. This example shows how different components of the azimuthal brightness function affect the shape of the images and the brightness profile of the ring.
Using the formula (\ref{finalvid}) for the cases when the coefficients are $a_m\neq0$ for $m=0,1,2$, we obtain the expression for the visibility function:
\begin{eqnarray}
V(u,\phi_u)=\pi a_0\int I_r(r)J_0(2\pi ur)rdr-\nonumber\\
-2\pi ia_1\cos(\phi_0-\phi_u)\int I_r(r)J_1(2\pi ur)rdr-\nonumber\\
-2\pi a_2\cos(2\phi_0-2\phi_u)\int I_r(r)J_2(2\pi ur)rdr.
\label{vid1}
\end{eqnarray}
This expression (\ref{vid1}) can be integrated explicitly if certain functions of the radial brightness of the source are specified.

If we assume that the photon's ring is an infinitely thin ring \cite{johnson2020}, then the radial brightness function of the ring will be proportional to the delta function, $I_r\sim\delta(r-r_0)$, where $r_0$ is the radius of the ring. Substituting this dependency into the expression for the visibility function (\ref{vid1}), we get
\begin{eqnarray}
V(u,\phi_u)=\pi a_0J_0(2\pi ur_0)r_0-2\pi ia_1\cos(\phi_0-\phi_u)J_1(2\pi ur_0)r_0-\nonumber\\
-2\pi a_2\cos(2\phi_0-2\phi_u)J_2(2\pi ur_0)r_0.
\label{vthinring}
\end{eqnarray}
In the case when the coefficient $a_2=0$, the phase of the visibility function is determined by the ratio of the Bessel functions of the first and zero order
\begin{eqnarray}
\tan\phi_v=\frac{ImV}{ReV}=-2\frac{a_1J_1(2\pi ur_0)}{a_0J_0(2\pi ur_0)}\cos(\phi_0-\phi_u).
\label{tanphidelta}
\end{eqnarray}
From this expression (\ref{tanphidelta}) it can be seen that the phase of the visibility function depends on the direction of projection of the base $\phi_u$ and on the direction of maximum brightness in the ring $\phi_0$. In particular, when the angle difference is $\phi_0-\phi\approx\frac{\pi}{2}$, the phase of the visibility function is close to zero.

In the case when the radial brightness in the ring is a uniform value $I_r=const$, the expression for the visibility function (\ref{vid1}) takes the following form \cite{chernov2021}
\begin{eqnarray}
V=\frac{a_0}{2}\left(\frac{b}{u}J_1(2\pi ub)-\frac{a}{u}J_1(2\pi ua)\right)-\nonumber\\
-i\pi a_1\cos(\phi_0-\phi_u)\bigg[\frac{b}{2u}(J_1(2\pi ub)H_0(2\pi ub)-J_0(2\pi ub)H_1(2\pi ub))-\nonumber\\
-\frac{a}{2u}(J_1(2\pi ua)H_0(2\pi ua)-J_0(2\pi ua)H_1(2\pi ua))\bigg]+\nonumber\\
+a_2\cos(2\phi_0-2\phi_u)\frac{2\pi ubJ_1(2\pi ub)-2\pi uaJ_1(2\pi ua)+2J_0(2\pi ub)-2J_0(2\pi ua)}{2\pi u^2}
\label{vidIconst}
\end{eqnarray}
where $H$ is the Struve function. This analytical expression takes into account a uniform brightness distribution and includes both real and complex components, which allows for more accurate modeling of the phase of the visibility function.

\begin{figure}
\centering
\includegraphics[width=1.2\linewidth]{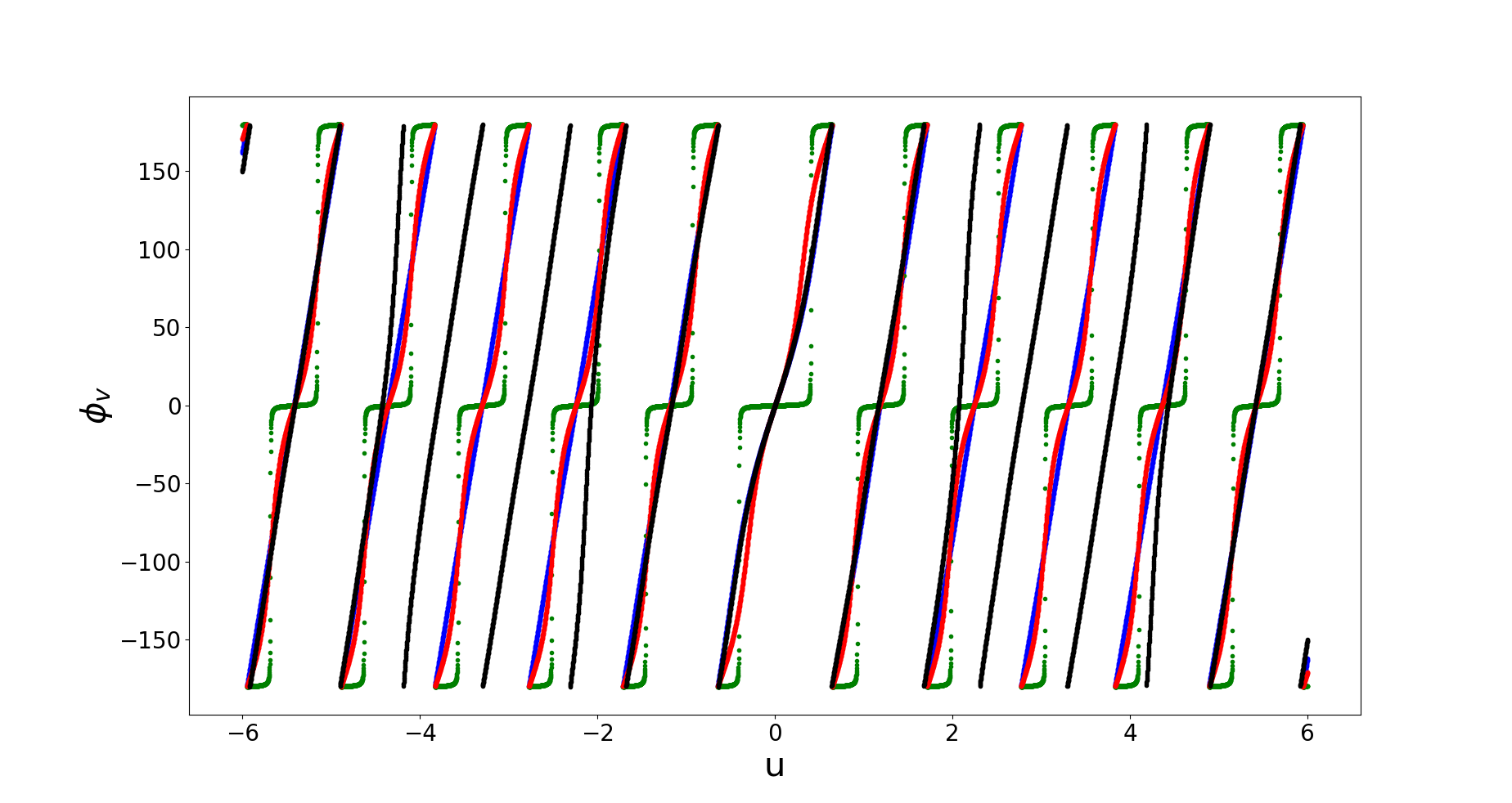}
\caption{The figure shows the dependence of the phase of the visibility function in degrees on the projection of the base in units of the Earth's diameter for thin and thick ring models with a uniform brightness distribution.}
\label{fig2}
\end{figure}

The figure (\ref{fig2}) shows the dependence of the phase of the visibility function in degrees on the projection of the base, expressed in units of the diameter of the Earth, in the direction of $\phi_u=0$. The maximum brightness in the ring is located at an angle of $\phi_0=\pi$. The following parameters were used for the models: the blue curve represents models of an infinitely thin ring (\ref{vthinring}) with a radius of $r_0=20$ microseconds of arc and coefficients $a_0=2$, $a_1=1$, $a_2=0$; the green curve represents models of an infinitely thin ring (\ref{vthinring}) with radius $r_0=20$ microseconds of arc and coefficients $a_0=2$, $a_1=0.01$, $a_2=0$; The red curve represents models of an infinitely thin ring (\ref{vthinring}) with a radius of $r_0=20$ microseconds of arc and coefficients $a_0=2$, $a_1=1$, $a_2=1$; the black curve represents models of a thick ring with a uniform radial brightness distribution (\ref{vidIconst}), with an outer and inner radius of $b=25$ and $a=15$ microseconds of arc, respectively, and with coefficients $a_0=2$, $a_1=a_2=1$. It can be seen from the figure (\ref{fig2}) that for small values of the projections of the bases ($u<1.5$), curves with the same coefficients $a_0$ and $a_1$ agree well with each other, despite the significant difference in the parameters of the rings. As the projection of the bases increases ($u>1.5$), the nature of the dependence changes: the phase of the visibility function for different models begins to differ significantly. This is due to the fact that with large projections of the bases, the components of the ring structure have a greater effect on the phase, making it more sensitive to specific geometric parameters.

\section{MHD models}

To compare the above theory with numerical observations, a three-dimensional MHD simulation of the accretion of an ideal fluid onto black holes was performed. The simulation was performed using the iharm3d\cite{gammie2003,noble2006} code. The value of the black hole's spin was set to $a=0.6$. For three slices at time points $t=0;1000;2000$, images of black hole were constructed using the ipole\cite{moscibrodzka2018,noble2007} code. Additional information about the parameters of MHD modeling and methods for constructing images of black holes can be found in \cite{chernov2021b,chernov2026b}. Images of black holes for the corresponding slices at frequencies of 230 and 690 GHz are shown in Figures (2), (4) and (6) of the article \cite{chernov2026b}. The amplitudes (modulus) of the visibility functions and their comparison with analytical models of thick and thin rings are shown in Figures (3), (5) and (7) of the same work \cite{chernov2026b}. This section will present the results of modeling the phase of the visibility function for these slices at a frequency of 230 GHz and their comparison with analytical models.

\begin{figure}
\centering
\includegraphics[width=1.0\linewidth]{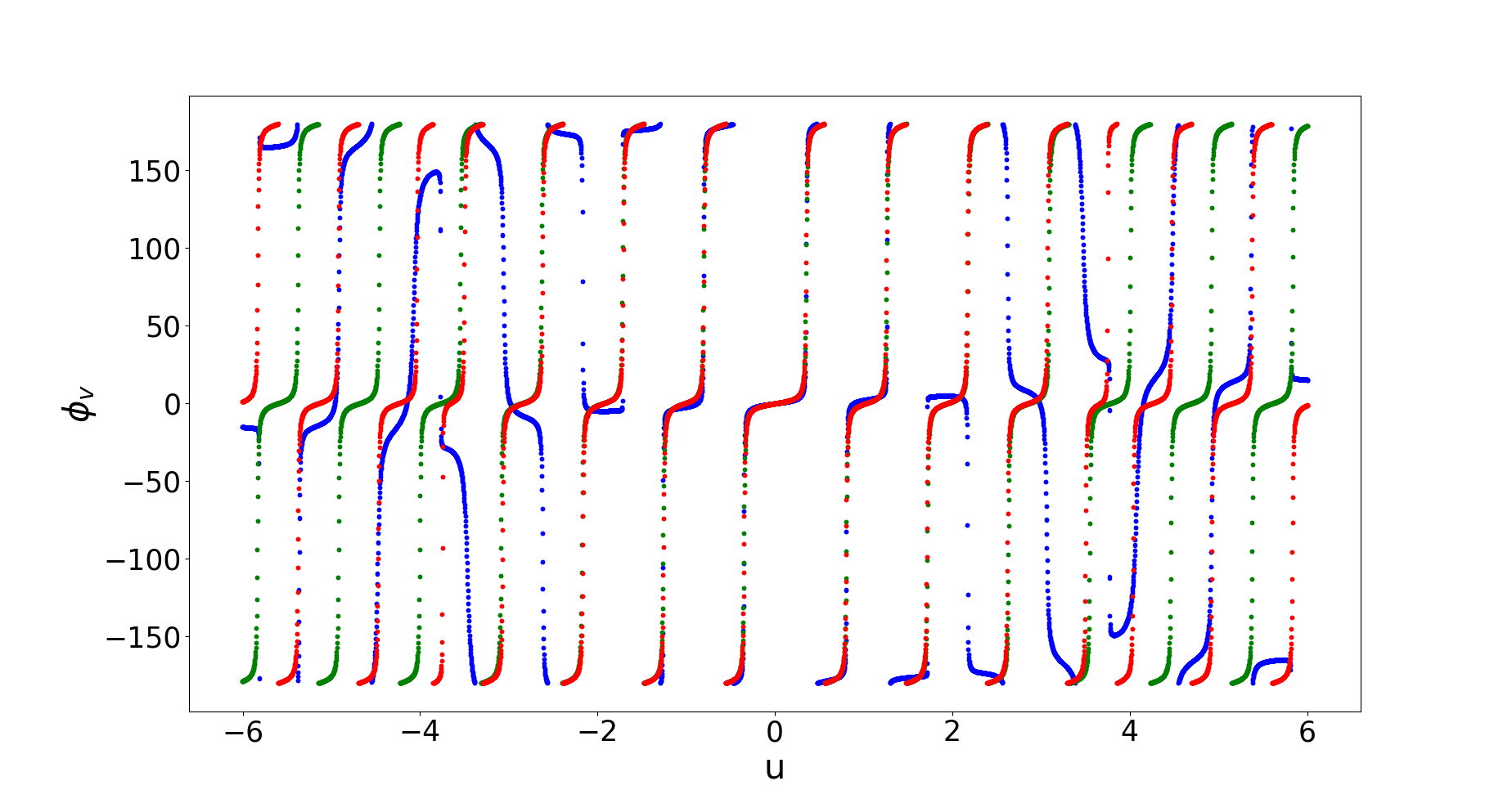}
\caption{The figure shows the dependence of the phase of the visibility function on the projection of the base in units of the Earth's diameter for the mhd model at $t=0$ and the corresponding analytical models of thin and thick disks.}
\label{fig3}
\end{figure}

The figure (\ref{fig3}) shows the dependence of the phase of the visibility function in degrees on the projection of the base, expressed in units of the diameter of the Earth. The blue curve corresponds to the image model of the black hole obtained by MHD modeling at the initial moment of time $t=0$, in the direction of $\phi_u=0$. The green curve corresponds to the model of an infinitely thin ring with a radius of $r_0=23.0428$ $\mu as$ and with Fourier expansion coefficients: $a_0=2$, $a_1=0.05$ and $a_2=0$. The red curve corresponds to the model of a thick ring with an inner and outer radius of $a=20.2183$, $b=25.8673$ microseconds of arc, respectively. The Fourier expansion coefficients for this model are the same as for the infinitely thin ring model: $a_0=2$, $a_1=0.05$, and $a_2=0$. The model parameters were selected by comparing the amplitude of the visibility function and are taken from \cite{chernov2026}.

\begin{figure}
\centering
\includegraphics[width=1.0\linewidth]{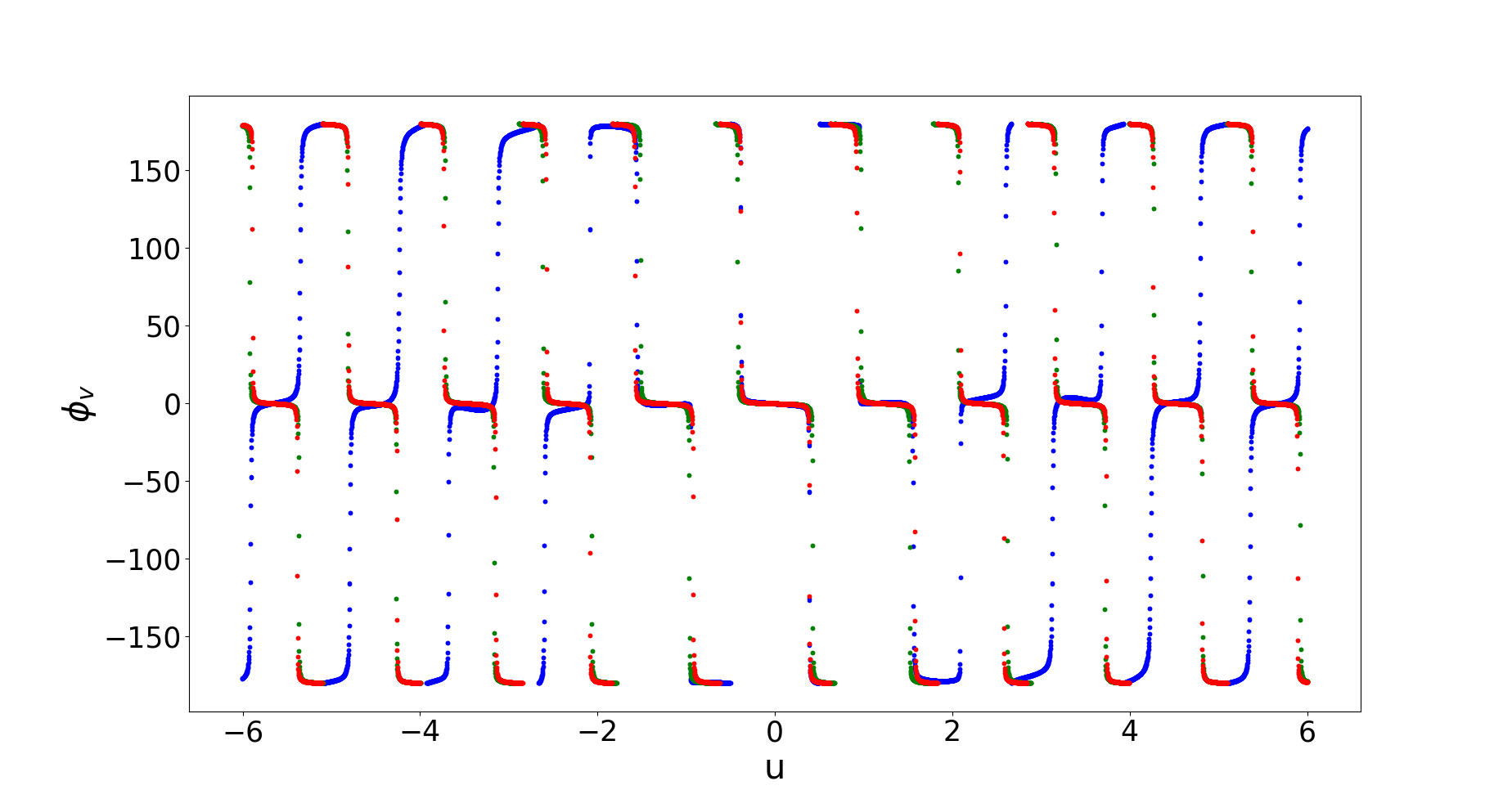}
\caption{The figure shows the dependence of the phase of the visibility function on the projection of the base in units of the Earth's diameter for the mhd model at $t=1000$ and the corresponding analytical models of thin and thick disks.}
\label{fig4}
\end{figure}

The figure (\ref{fig4}) shows the dependence of the phase of the visibility function in degrees on the projection of the base, expressed in units of the diameter of the Earth. The blue curve corresponds to the image model of the black hole obtained by MHD modeling for the slice at time $t=1000$, in the direction of $\phi_u=0$. The green curve corresponds to the model of an infinitely thin ring with a radius of $r_0=19.15$ microseconds of arc and with Fourier expansion coefficients equal to: $a_0=-2$, $a_1=-0.01$ and $a_2=0$. The red curve corresponds to two thick ring models with parameters taken from table (2) of the work (\cite{chernov2026b}) $a_1=17.8$, $b_1=20.5$, $c_1=0.011$ and $a_2=12.0153$, $b_2=30.8804$, $c_2=0.0023$ microseconds of arc. The Fourier expansion coefficients for the thick ring model are the same as for the infinitely thin ring model: $a_0=-2$, $a_1=-0.01$, and $a_2=0$.

\begin{figure}
\centering
\includegraphics[width=1.0\linewidth]{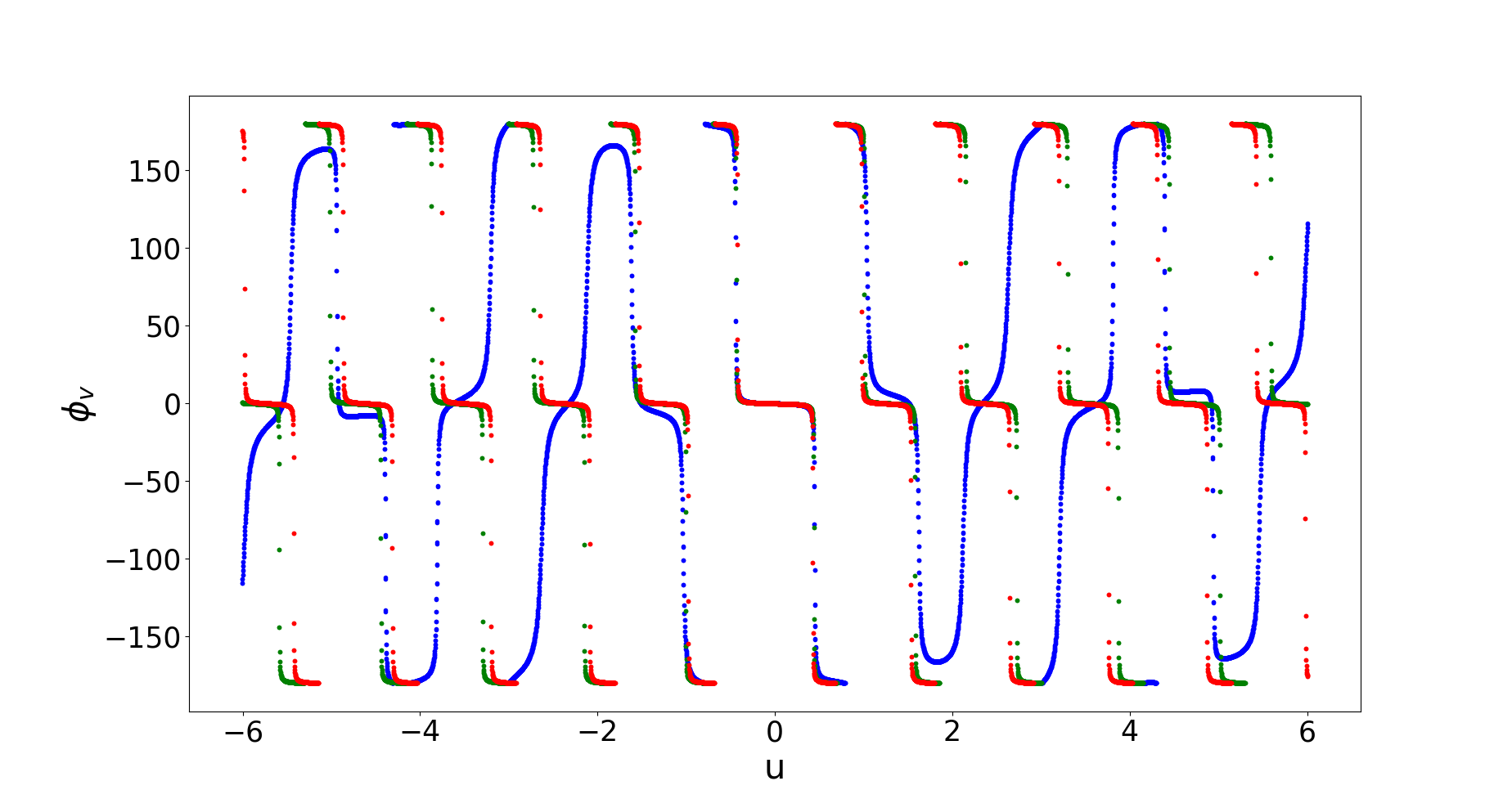}
\caption{The figure shows the dependence of the phase of the visibility function on the projection of the base in units of the Earth's diameter for the mhd model at $t=2000$ and the corresponding analytical models of thin and thick disks.}
\label{fig5}
\end{figure}

The figure (\ref{fig5}) shows the dependence of the phase of the visibility function in degrees on the projection of the base, expressed in units of the diameter of the Earth. The blue curve corresponds to the image model of the black hole obtained by MHD modeling for a slice at time $t=2000$, in the direction of $\phi_u=0$. The green curve corresponds to the model of an infinitely thin ring with a radius of $r_0=18.4$ microseconds of arc and Fourier expansion coefficients: $a_0=-2$, $a_1=-0.01$ and $a_2=0$. The red curve corresponds to the thick ring model with parameters taken from table (3) of the work (\cite{chernov2026b}) $a_1=10.9259$, $b_1=25.1576$, $c_1=0.0011$, $a_2=17.7223$, $b_2=20.1513$, $c_2=0.0049$ and $a_3=$12.2192, $b_3=$19.0090, $c_3=0.0014$ microseconds of arc. The Fourier expansion coefficients for the thick ring model are the same as for the infinitely thin ring model: $a_0=-2$, $a_1=-0.01$, and $a_2=0$.

From Figures (\ref{fig3}), (\ref{fig4}) and (\ref{fig5}) it can be seen that for small values of the projection bases ($|u|<2$ in units of the Earth's diameter), there is a good agreement between theoretical predictions and the results of numerical MHD models. Large values of the base projections are characterized by either a good degree of agreement or agreement with theoretical calculations, taking into account possible changes in the phase sign.

\section{Observations}

On April 14-15, 2018, a group of GMVA radio telescopes (Global Millimetre VLBI Array) made observations of a supermassive black hole in the elliptical galaxy M87 at a wavelength of $\lambda=3.5$ mm \cite{lu2023}. As a result, an image of a ring-shaped bright formation was obtained, in the center of which there is an area with reduced brightness. Such an image is interpreted as an image of a black hole. The diameter of the ring-shaped structure at a wavelength of 3.5 mm turned out to be about 50 percent larger than when observed at a wavelength of 1.3 mm, and is about $64$ microseconds of arc\cite{lu2023}. The radiation flux at this wavelength is estimated at about $J_{86}\approx0.5-0.6$ Yang\cite{lu2023}.

\begin{figure}
\centering
\includegraphics[width=1.1\linewidth]{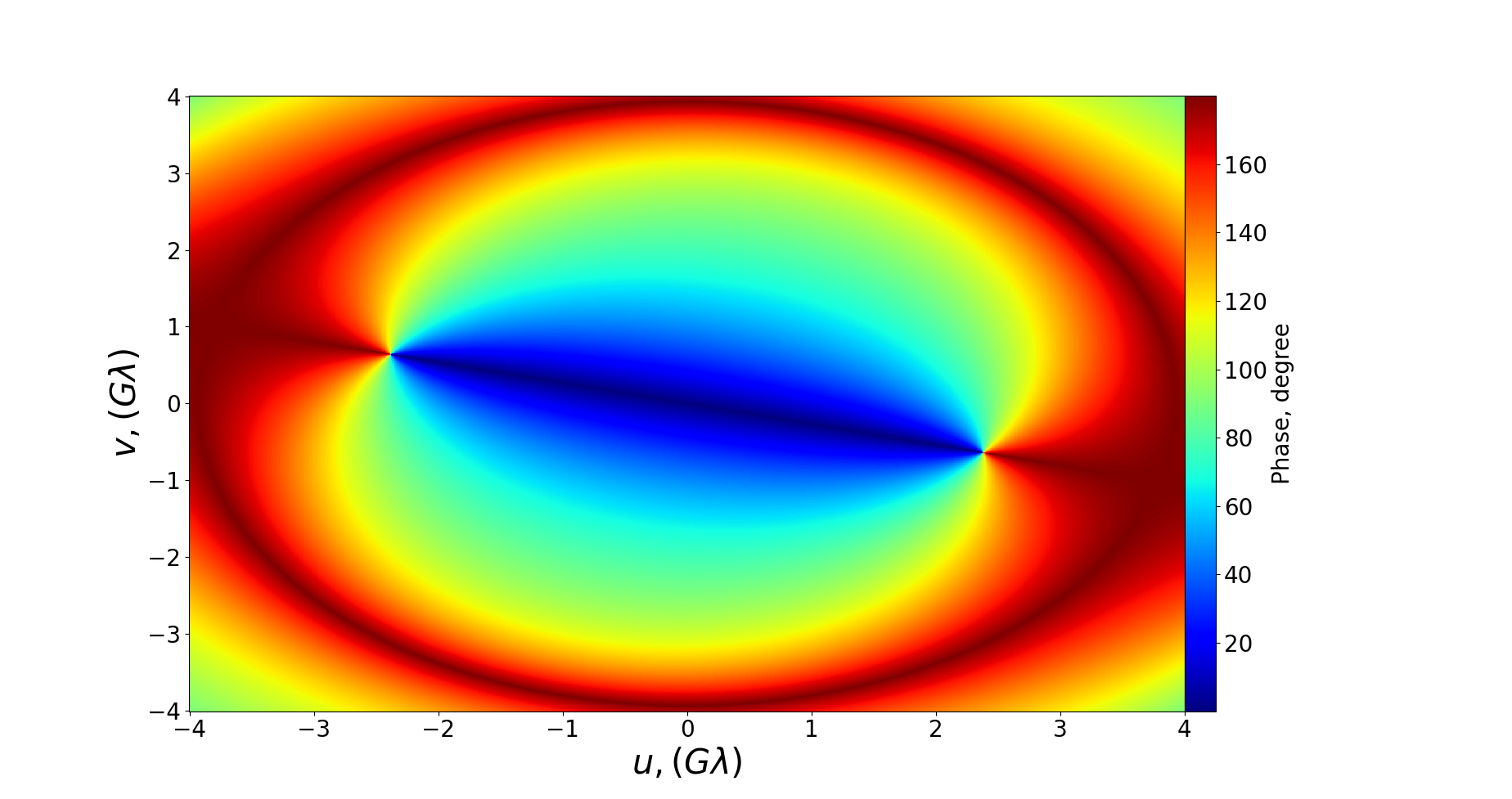}
\caption{The figure shows a two-dimensional modulus of the phase of the visibility function, calculated using the formula (\ref{tanphidelta}) and given in degrees. The projections of the base are shown in wavelength units, according to Figure s2 of the article \cite{lu2023}.}
\label{fig6}
\end{figure}

Figure s2 in the additional materials of the article \cite{lu2023} shows the phase module of the visibility function on the uv plane. It can be seen from this figure that for small projections of the bases (i.e., for |u|,|v|<2G$\lambda$), the phase modulus of the visibility function is approximately zero. While with large projections of bases (|u|,|v|>2G$\lambda$), the phase modulus of the visibility function tends to 180 degrees. The figure (\ref{fig6}) shows the dependence of the phase modulus of the visibility function on the uv plane calculated using the formula (\ref{tanphidelta}). The direction corresponding to the maximum brightness was chosen to be $\phi_0=7\pi/12$ radians, and the radius of the ring was set to $r_0=32$ microseconds of arc. From this figure, it can be seen that inside small projections of bases (|u|,|v|<2G$\lambda$), the phase modulus of the visibility function is close to zero, and for large projections of bases (|u|,|v|>2G$\lambda$), it is close to 180 degrees. However, it is not possible to accurately compare the theoretical curve of the phase modulus of the visibility function with the observational data due to insufficient filling of the uv plane.

Figure 31-7 of the collection of articles \cite{taylor1999} shows the amplitude and phase of the visibility function depending on the projection of the base of radio interferometric observations of Venus at a wavelength of 2 cm. obtained using VLA. The observational data are in good agreement with the theoretical relationship presented in (\ref{tanphidelta}) and shown in Fig. (\ref{fig2}).

\section{Conclusions}

The present work is devoted to the study of the phase of the visibility function in radio interferometry with space bases. This feature is a fundamental tool for obtaining highly detailed images of supermassive black holes. The analysis showed that the mathematical model of the visibility function, divided into a real and imaginary part, allows you to accurately relate the phase of the visibility function to the source parameters. When studying the conditions of small projections of bases, universal relations were derived that relate the phase of the visibility function to the geometric and physical characteristics of the system.

Three-dimensional numerical magnetohydrodynamic modeling was performed to verify the theoretical model. As part of these studies, the phases of the visibility function were constructed depending on the projection of bases up to six Earth diameters. The results obtained show a good agreement between theoretical predictions and MHD models with small projections of the bases. 

Also, the simplest theoretical model was compared with the observational data conducted by the GMVA group of radio telescopes at a frequency of 86 GHz for the radio source M$87^*$. A qualitative correspondence was shown between the phase module of the visibility function depending on the projection of the base. These results can be directly applied to solving the phase closure problem.

\section{Appendix}

To calculate the integrals in the expression (\ref{vidIconst}), the following tabular formulas are used \cite{gradshtein}:
\begin{eqnarray}
\int xJ_0(2\pi ux)dx=\frac{x}{2\pi u}J_1(2\pi ux),\nonumber\\
\int xJ_1(2\pi ux)dx=\frac{x}{4u}[J_1(2\pi ux)H_0(2\pi ux)-J_0(2\pi ux)H_1(2\pi ux)],\nonumber\\
\int xJ_2(2\pi ux)dx=\frac{-1}{(2\pi u)^2}(2J_0(2\pi ux)+2\pi uxJ_1(2\pi ux)),
\end{eqnarray}
where $H$ is the Struve function.

\end{document}